\documentclass[pra,aps,showpacs,twocolumn]{revtex4}
\usepackage{amssymb}
\usepackage{graphicx}
\usepackage{latexsym}
\usepackage{amsmath}
\usepackage[dvipdfm]{hyperref}

\begin{document}

\title{coherence of quantum Gaussian channels}
\author{Jianwei Xu}
\email{xxujianwei@nwafu.edu.cn}
\affiliation{College of Science, Northwest A\&F University, Yangling, Shaanxi 712100,
China}
\date{\today }

\begin{abstract}
Coherence is a basic notion for quantum states. Instead of quantum states,
in this work, We establish a resource theory for quantifying the coherence
of Gaussian channels. To do this, we propose the definitions of incoherent
Gaussian channels and incoherent Gaussian superchannels.
\end{abstract}

\pacs{03.65.Ud, 03.67.Mn, 03.65.Aa}
\maketitle

\section{Introduction}

Coherence is a fundamental ingredient for quantum physics and an important
resource for quantum information science. In the past few years there have
been many works focusing on the coherence of quantum states and have
achieved fruitful results both in theories and applications (recent reviews
see \cite{Plenio-2016-RMP,Hu2018-PhysicsReports} and references therein).
However, the past research for coherence is only for quantum states. Quantum
information processings inevitably involve quantum channels, then it is
necessary to consider the coherence of quantum channels. From a higher
perspective, quantum states can be regarded as degenerated quantum channels
\cite{Gour-2019-IEEE}, or we call a resource possessed in quantum states as
static resource while a resource possessed in quantum channels as dynamic
resource \cite{Winter-IEEE-2008}. Recently many researchers begin to exploit
the coherence of quantum channels \cite%
{Winter-2017-PRA,Korzekwa-2018-NJP,Datta-2018-AOP,Plenio-2019-PRL,
LiuYuan-2019,LiuWinter-2019,Xu-2019-arXiv}.

All these research about coherence both for quantum states or quantum
channels mainly aims to the finite-dimensional systems setting. When we
consider the case of infinite dimensional systems, we will encounter many
difficulties. Firstly, a physical quantum state in an infinite-dimensional
system should be convergent in some sense. Furthermore, the expressions for
finite-dimensional systems often become hard to evaluate for
infinite-dimensional systems even they are convergent such as the relative entropy of coherence for quantum states \cite{BCP-2014-PRL}. Quantum Gaussian
states are a class of infinite-dimensional states which play significant
role in quantum optics and in quantum information theory (reviews see e.g.
\cite{Lloyd-2012-RMP,Serafini-2017-book}) with thermal states, coherent
states and squeezed states as special cases of Gaussian states. Till now the
coherence of quantum Gaussian states has been discussed in many works e.g.
\cite{Fan-2016-PRA,Xu-2016-PRA,Zheng-2016-PRA,Illuminati-2016-arxiv,Jeong-2017-PRL,Paris-2017-PRA}.

In this work, we turn to study the coherence of Gaussian channels under the
quantum resource theory (QRT) \cite{ChitambarGour-2019-RMP}. This paper is
organized as follows. In section II, we give the definitions of incoherent
Gaussian channels and incoherent Gaussian superchannels, we also explore the
structures of them. In section III, we establish a QRT for quantifying the
coherence of Gaussian channels, and explicitly provide such a measure.
Section VI is a brief summary. For clarity of the structure and easy
reading, we postpone most necessary proofs for this work to the Appendix
part.
\section{Incoherent Gaussian channels and incoherent Gaussian superchannels}

Let $\{|j\rangle \}_{j=0}^{\infty }$ be an orthonormal basis of Hilbert
space $H,$ we call the basis $\{|j\rangle \}_{j=0}^{\infty }$ Fock basis and
call each state $|j\rangle $ Fock state. Note that the notion of coherence
is dependent on the chosen orthonormal basis, so we always suppose the Fock
basis is fixed, or call it reference basis. When we consider the $n$-fold
tensor Hilbert space $H^{\otimes n},$ we adopt the tensor basis $%
(\{|j\rangle \}_{j=0}^{\infty })^{\otimes n}$ as its reference basis.

For a state $\rho $ on $H^{\otimes n},$ its characteristic function $\chi
(\rho ,\lambda )$ is defined as
\begin{eqnarray}
\chi (\rho ,\lambda )&=&tr[\rho D(\lambda )],  \label{eq1} \\
D(\lambda )&=&\Pi _{j=1}^{n}D(\lambda _{j}),  \label{eq2} \\
D(\lambda _{j})&=&\exp (\lambda _{j}a_{j}^{\dag }-\lambda _{j}^{\ast }a_{j}),  \label{eq3}
\end{eqnarray}
where $\lambda =(\lambda _{1x},\lambda _{1y},\lambda _{2x},\lambda
_{2y},...,\lambda _{nx},\lambda _{ny})^{t}\in R^{2n}$, $\lambda _{j}=\lambda
_{jx}+i\lambda _{y}$, $\lambda _{jx}$, $\lambda _{y}\in R$, $%
a_{j},a_{j}^{\dag }$ are the annihilation and creation operators acting on
the $j$th Hilbert space $H,$ and *, $t$, \dag\ represent the complex
conjugation, transposition and Hermitian conjugation. Notice that we use
both notations $\lambda _{j}=(\lambda _{jx},\lambda _{jy})$ $\ $and $\lambda
_{j}=\lambda _{jx}+i\lambda _{y}$ according to different contexts.

A state $\rho $ on $H^{\otimes n}$ is called a Gaussian state if its
characteristic function $\chi (\rho ,\lambda )$ has the form
\begin{eqnarray}
\chi (\rho ,\lambda )=\exp [-\frac{1}{2}\lambda ^{t}\Omega V\Omega
^{t}\lambda -i(\Omega d_{0})^{t}\lambda ],  \label{eq4}
\end{eqnarray}
where $d_{0}=(d_{01x},d_{01y},d_{02x},d_{02y},...,d_{0nx},d_{0ny})^{t}\in
R^{2n}$ is called displacement vector, $d_{0j}=d_{0jx}+id_{0jy}$, $d_{0jx}$,
$d_{0jy}\in R$, $\Omega =\oplus _{j=1}^{n}\omega $, $\omega =\left(
\begin{array}{cc}
0 & 1 \\
-1 & 0%
\end{array}%
\right) $, $V$ is a $2n\times 2n$ real symmetric matrix called covariance
matrix satisfying the uncertainty relation \cite{Simon-1994-PRA}
\begin{eqnarray}
V+i\Omega \geq 0. \label{eq5}
\end{eqnarray}
We see that a Gaussian state $\rho $ is completely described by $(V,d_{0})$,
then we denote $\rho =\rho (V,d_{0})$, and also we denote the set of all
Gaussian states on $H^{\otimes n}$ by $\mathcal{GS}_{n}.$ Notice that $%
\mathcal{GS}_{n}$ is not a convex set. \newline

\textbf{Definition 1.} We call a Gaussian state incoherent Gaussian state if
it is a thermal state. \newline

This definition comes from the fact that the incoherent state is defined as
the diagonal state in reference basis \cite{BCP-2014-PRL}, and the diagonal
Gaussian states are just the thermal states (an explicit proof see \cite%
{Xu-2016-PRA}). We call the set of all incoherent Gaussian states as $%
\mathcal{IGS}_{n}$, that is
\begin{eqnarray}
\mathcal{IGS}_{n}=\{\otimes _{j=1}^{n}\rho _{th}(\nu _{j})|\nu _{j}\geq 1 \
\forall \ j \},  \label{eq6} \\
\rho _{th}(\nu _{j})=\frac{2}{\nu _{j}+1}\sum_{k=0}^{\infty}(\frac{\nu _{j}-1%
}{\nu _{j}+1})^{k}|k\rangle \langle k|. \label{eq7}
\end{eqnarray}

A quantum channel is defined as a linear map from quantum states into
quantum states with the conditions of complete positivity and trace
preservation \cite{Nielsen-2000-book}. Now we say that a quantum channel is
Gaussian when it transforms Gaussian states into Gaussian states. A Gaussian
channel $\phi $ on $\mathcal{GS}_{n}$ can be described by $\phi (T,N,d)$ it
acts on $\rho (V,d_{0})\in \mathcal{GS}_{n}$ as \cite{Holevo-1999-PRA}
\begin{eqnarray}
&&d_{0}\rightarrow Td_{0}+d, \label{eq8} \\
&&V\rightarrow TVT^{t}+N, \label{eq9}
\end{eqnarray}
where $d\in R^{2n}$ is a displacement vector, while $T$ and $N=N^{t}$ are $%
2n\times 2n$ real matrices, which must satisfy the complete positivity
condition
\begin{eqnarray}
N+i\Omega -iT\Omega T^{t}\geq 0. \label{eq10}
\end{eqnarray}
When $N=0$, and $T$ is a symplectic matrix, i.e.
\begin{eqnarray}
T\Omega T^{t}=\Omega, \label{eq11}
\end{eqnarray}
we call such channel unitary Gaussian channel.

We denote the set of all Gaussian channels $\phi (T,N,d)$ on $\mathcal{GS}%
_{n}$ by $\mathcal{GC}_{n}.$

\bigskip

\textbf{Definition 2.} A Gaussian channel $\phi \in \mathcal{GC}_{n}$ is
called incoherent, if $\phi (\rho _{th})\in \mathcal{IGS}_{n}$ for $\forall $
$\rho _{th}\in \mathcal{IGS}_{n}.$

\bigskip

We denote the set of all incoherent Gaussian channels by $\mathcal{IGC}_{n}.$

We now determine the structure of $\mathcal{IGC}_{n}.$

\bigskip

\textbf{Theorem 1.} A Gaussian channel $\phi (T,N,d)\in \mathcal{GC}_{n}$ is
incoherent iff (if and only if)
\begin{eqnarray}
d&=&0,  \label{eq12} \\
T&=&\{t_{j}T_{j}\}_{j=1}^{n}\in \mathcal{T}_{n},  \label{eq13} \\
N&=&\oplus _{j=1}^{n}\omega _{j}I_{2},  \label{eq14} \\
\omega _{j}&\geq& |1-\sum_{k:r(k)=j}t_{k}^{2}\det T_{k}|,\forall \ j, \label{eq15}
\end{eqnarray}
where $t_{j}$, $\omega _{j}\in R$, $T_{j}$ is a $2\times 2$ real matrix for $%
\forall $ $j$; $I_{2}$ is the $2\times 2$ identity; $\mathcal{T}_{n}$
denotes the set of all $2n\times 2n$ real matrices such that for any matrix $%
T\in \mathcal{T}_{n},$  the $(2j-1,2j)$\ two columns have just one $2\times
2 $ real matrix $t_{j}T_{j}$ with $T_{j}T_{j}^{t}=I_{2},$ i.e., orthogonal,
located in $(2r(j)-1,2r(j))$ rows for $\forall $ $j$, $r(j)\in
\{k\}_{k=1}^{n}$, and other elements are all zero. \newline

Note that the identity Gaussian channel $\phi (T=I_{2n},N=0,d=0)\in \mathcal{%
IGC}_{n},$ where $I_{2n}$ is the $2n\times 2n$ identity.

Superchannel is a completely positive linear map transforming channels into
channels \cite{Chiribella-2008-EPL,Gour-2019-IEEE}. Gaussian superchannel is
hence defined as superchannel transforming Gaussian channels into Gaussian
channels. Let $\mathcal{GSC}_{n} $ denote the set of all Gaussian
superchannels acting on $\mathcal{GC}_{n}.$ For $\phi (T,N,d)\in \mathcal{GC}%
_{n}$, consider the Gaussian state $\rho _{\phi }(V,d_{0})\in \mathcal{GS}%
_{2n}$ with
\begin{eqnarray}
V&=&\left(
\begin{array}{cc}
TT^{t}ch2r+N & T\Sigma _{n}sh2r \\
\Sigma _{n}T^{t}sh2r & I_{2n}ch2r%
\end{array}%
\right),  \label{eq16} \\
d_{0}&=&\binom{d}{0},  \label{eq17} \\
\Sigma _{n}&=&\oplus _{j=1}^{n}\left(
\begin{array}{cc}
1 & 0 \\
0 & -1%
\end{array}%
\right),  \label{eq18}
\end{eqnarray}
and $r\in R,$ $ch2r=\frac{e^{2r}+e^{-2r}}{2}$, $sh2r=\frac{e^{2r}-e^{-2r}}{2}%
.$ The fact that $\rho _{\phi }$ is indeed a Gaussian state see e.g. section
5.5.2 in \cite{Serafini-2017-book}. $\rho _{\phi }$ is called the Choi state
of Gaussian channel $\phi .$ From this correspondence between $\phi $ and $%
\rho _{\phi }$, we provide a characterization of $\mathcal{GSC}_{n}.$ The
idea is as follows. Suppose $\Phi \in \mathcal{GSC}_{n},$ which transforms $%
\phi \in \mathcal{GC}_{n}$ as $\Phi (\phi )\in \mathcal{GC}_{n}$, then we
will find a Gaussian channel which transforms $\rho _{\phi }$ into $\rho
_{\Phi (\phi )}$. The result is Theorem 2 below.

\bigskip

\textbf{Theorem 2.} A Gaussian superchannel $\Phi \in \mathcal{GSC}_{n}$ can
be represented by $\Phi (A,O,Y,\overline{d}),$ and for $\phi (T,N,d)\in
\mathcal{GC}_{n},$ we have $\Phi \lbrack \phi (T,N,d)]=\psi (T^{\prime
},N^{\prime },d^{\prime })$ with
\begin{eqnarray}
T^{\prime }&=&AT\Sigma _{n}O^{t}\Sigma _{n},  \label{eq19} \\
N^{\prime }&=&ANA^{t}+Y,  \label{eq20} \\
d^{\prime }&=&Ad+\overline{d}, \label{eq21}
\end{eqnarray}
where $A,O,Y$ are all $2n\times 2n$ real matrices, $Y=Y^{t},$ $OO^{t}=I_{2n}$%
, $\overline{d}\in R^{2n}$, and
\begin{eqnarray}
Y+i\Omega -iA\Omega A^{t}\geq 0,  \label{eq22} \\
i\Omega -iO\Omega O^{t}\geq 0. \label{eq23}
\end{eqnarray}

\bigskip

We can also express any Gaussian superchannel in terms of compositions of Gaussian channels, this is Theorem 3 below.

\bigskip

\textbf{Theorem 3.} A Gaussian superchannel $\Phi (A,O,Y,\overline{d})$ can
be represented by $\Phi (\phi )=\phi _{2}\circ \phi \circ \phi _{1}$ for $%
\forall $ $\phi \in \mathcal{GC}_{n}$ with fixed $\phi
_{1}(T_{1},N_{1},d_{1})$, $\phi _{2}(T_{2},N_{2},d_{2})\in \mathcal{GC}_{n}$%
. One such representation is
\begin{eqnarray}
&&T_{1}=\Sigma _{n}O^{t}\Sigma _{n},N_{1}=0,d_{1}=0;  \label{eq24} \\
&&T_{2}=A,N_{2}=Y,d_{2}=\overline{d}.  \label{eq25}
\end{eqnarray}

\bigskip

To establish a QRT for coherence of Gaussian channels, we need to specify
the definition of incoherent Gaussian superchannel.

\bigskip

\textbf{Definition 3.} A Gaussian superchannel $\Phi \in \mathcal{GSC}_{n}$
is called incoherent if $\Phi (\phi )\in \mathcal{IGC}_{n}$ for $\forall $ $%
\phi \in \mathcal{IGC}_{n}.$

\bigskip

We denote the set of all incoherent Gaussian superchannels by $\mathcal{IGSC}%
_{n}.$ The structure of $\mathcal{IGSC}_{n}$ is as following Theorem 4.

\bigskip

\textbf{Theorem 4.} For Gaussian superchannel $\Phi (A,O,Y,\overline{d}),$
the following (1), (2) and (3) are equivalent. \newline
(1). $\Phi (A,O,Y,\overline{d})$ is incoherent. \newline
(2). $\overline{d}=0$, $A\in \mathcal{T}_{n}$, $O\in \mathcal{T}_{n}$, $%
Y=\oplus _{j=1}^{n}\eta _{j}I_{2}$, $\eta _{j}\in R.$ \newline
(3). $\Phi (\phi )=\chi _{2}\circ \phi \circ \chi _{1}$ for $\forall $ $\phi
\in \mathcal{GC}_{n}$, with fixed $\chi _{1}$, $\chi _{2}\in \mathcal{IGC}%
_{n}$.

\bigskip

\section{Framework for quantifying coherence of Gaussian channels.}

With the definitions of incoherent Gaussian channels and incoherent Gaussian
superchannels, we now establish a quantum resource theory for quantifying
the coherence of Gaussian channels.

We propose the following conditions that any coherence measure for Gaussian
channels should satisfy. \newline
(C1). $C(\phi )\geq 0$ $\forall $ $\phi \in \mathcal{GC}_{n},$ and $C(\phi
)=0$ iff $\phi \in \mathcal{IGC}_{n}.$ \newline
(C2). $C[\Psi (\phi )]\leq C(\phi ),$ $\forall $ $\phi \in \mathcal{GC}_{n},$
$\forall $ $\Psi \in \mathcal{IGSC}_{n}.$

Note that, (C2) is equivalent to (C3a)+(C3b) below. This can be seen by
Theorem 4 and the fact that the identity Gaussian channel $\phi
(T=I_{2n},N=0,d=0)\in \mathcal{IGC}_{n}.$ \newline
(C3a). $C(\phi \circ \chi _{1})\leq C(\phi ),\forall $ $\phi \in \mathcal{GC}%
_{n},\chi _{1}\in \mathcal{IGC}_{n}.$ \newline
(C3b). $C(\chi _{2}\circ \phi )\leq C(\phi ),\forall $ $\phi \in \mathcal{GC}%
_{n},\chi _{2}\in \mathcal{IGC}_{n}.$

Theorem 5 below provides a way to construct a class of coherence measures
for Gaussian channels, the proof is simple.

\bigskip

\textbf{Theorem 5.} If the functional $C:\mathcal{GS}_{n}\rightarrow R$
satisfies (B1). $C(\rho )\geq 0$ $\forall $ $\rho \in \mathcal{GS}_{n},$ and
$C(\rho )=0$ iff $\rho \in \mathcal{IGS}_{n};$ (B2). $C[\chi (\rho )]\leq
C(\rho ),$ $\forall $ $\rho \in \mathcal{GS}_{n},$ $\forall $ $\chi \in
\mathcal{IGC}_{n},$ \newline
then
\begin{eqnarray}
C(\phi )=\sup_{\rho _{th}\in \mathcal{IGS}_{n}}C[\phi (\rho _{th})]  \label{eq26}
\end{eqnarray}
is a coherence measure for Gaussian channels.

\bigskip

Consider the functional $C_{r}:\mathcal{GS}_{n}\rightarrow R$,
\begin{eqnarray}
C_{r}(\rho )=\inf_{\sigma \in \mathcal{IGS}_{n}}S(\rho ||\sigma ),\ \forall
\ \rho \in \mathcal{GS}_{n},  \label{eq27}
\end{eqnarray}
with $S(\rho ||\sigma )=tr(\rho \log _{2}\rho )-tr(\rho \log _{2}\sigma )$ the
relative entropy. It is easy to check that $C_{r}$ satisfies (B1) and (B2),
and also $C_{r}(\rho )$ has the analytical expression as \cite{Xu-2016-PRA}
\begin{eqnarray}
&&C_{r}[\rho (V,d_{0})]=\sum_{j=1}^{n}f(\overline{n_{j}})-\sum_{j=1}^{n}f(%
\frac{\nu _{j}-1}{2}),  \label{eq28} \\
&&f(x)=(x+1)\log _{2}(x+1)-x\log _{2}x,\ x\in R,   \label{eq29} \\
&&\overline{n_{j}}=\frac{1}{4}%
[V_{11}^{(j)}+V_{22}^{(j)}+d_{0jx}^{2}+d_{0jy}^{2}-2],  \label{eq30}
\end{eqnarray}%
where $S(\rho )=$ $\sum_{j=1}^{n}f(\frac{\nu _{j}-1}{2})\ $ is the entropy
of $\rho $ \cite{Holevo-1999-PRA}, $\{\nu _{j}\}_{j=1}^{n}$ are the
symplectic eigenvalues of $V$ \cite{Lloyd-2012-RMP}, $\overline{n_{j}}$ is
determined by the $j$th-mode covariance matrix $V^{(j)}$ and displacement
vector $d_{0j}$.

Consequently,
\begin{eqnarray}
C_{r}(\phi )=\sup_{\rho _{th}\in \mathcal{IGS}_{n}}C_{r}[\phi (\rho _{th})]   \label{eq31}
\end{eqnarray}
is a coherence measure for Gaussian channels.

It is straightforward to check that $C_{r}$ satisfies (B3) and (B4) below.
\newline
(B3). $C_{r}(\rho _{1}\otimes \rho _{2})=C_{r}(\rho _{1})+C_{r}(\rho
_{2}),\forall $ $\rho _{1}\in \mathcal{GS}_{n_{1}},\rho _{2}\in \mathcal{GS}%
_{n_{2}}.$ \newline
(C4). $C_{r}(\phi _{1}\otimes \phi _{2})=C_{r}(\phi _{1})+C_{r}(\phi
_{2}),\forall $ $\phi _{1}\in \mathcal{GC}_{n_{1}},\phi _{2}\in \mathcal{GC}%
_{n_{2}}.$

We give two concrete examples to show the calculation of coherence $%
C_{r}(\phi )$.

\bigskip

\textbf{Definition 4.} We call a Gaussian channel $\phi (T,N,d)\in \mathcal{%
GC}_{n}$ a constant Gaussian channel if $\phi (\rho )=\rho ^{\prime }$ for $%
\forall $ $\rho (V,d_{0})\in \mathcal{GS}_{n}$ where $\rho ^{\prime
}(V^{\prime },d_{0}^{\prime })\in \mathcal{GS}_{n}$ is fixed. Such constant
Gaussian channel can be represented as $\phi (T,N,d)=\phi (T=0,N=V^{\prime
},d=d_{0}^{\prime }).$

\bigskip

\emph{Exmple 1.} For a constant Gaussian channel, according to Eq. (\ref{eq31}),
\begin{eqnarray}
C_{r}[\phi (T=0,N=V^{\prime },d=d_{0}^{\prime })]=C_{r}[\rho ^{\prime
}(V^{\prime },d_{0}^{\prime })].   \label{eq32}
\end{eqnarray}

\emph{Exmple 2.} Coherence of displacement channels.

The displacement operator $D(\lambda )=\Pi _{j=1}^{n}D(\lambda _{j})$ in Eqs. (\ref{eq2}, \ref{eq3})
is a unitary Gaussian channel acting on the state $\rho$ as $D(\lambda
)\rho D(-\lambda )$. For $\rho (V,d_{0})\in \mathcal{GS}_{n},$ we have $%
D(\lambda )\rho (V,d_{0})D(-\lambda )\in \mathcal{GS}_{n}$, and its
characteristic function is
\begin{eqnarray}
&&\chi (D(\lambda )\rho (V,d_{0})D(-\lambda ),\mu )  \notag \\
&=&tr[D(\lambda )\rho (V,d_{0})D(-\lambda )D(\mu )]  \notag \\
&=&tr[\rho (V,d_{0})D(-\lambda )D(\mu )D(\lambda )]  \notag \\
&=&\exp (\mu \lambda ^{\ast }-\mu ^{\ast }\lambda )tr[\rho (V,d_{0})D(\mu )]
\notag \\
&=&tr[\rho (V,d_{0}+2\lambda )D(\mu )]  \notag \\
&=&\chi (\rho (V,d_{0}+2\lambda ),\mu ),   \label{eq33}
\end{eqnarray}
where we have used
\begin{eqnarray}
D(-\lambda )D(\mu )D(\lambda )=D(\mu )\exp (\mu \lambda ^{\ast }-\mu ^{\ast
}\lambda ).   \label{eq34}
\end{eqnarray}
and Eq. (\ref{eq4}).
Thus the Gaussian channel of displacement operator $D(\lambda )$ can be
written as
\begin{eqnarray}
&&\phi (T=I_{2n},N=0,d=2\lambda )  \notag \\
&=&\otimes _{j=1}^{n}\phi (T=I_{2},N=0,d_{j}=2\lambda _{j}).   \label{eq35}
\end{eqnarray}
From (C4) and Eqs. (\ref{eq31}, \ref{eq28}) we have
\begin{eqnarray}
&&C_{r}[D(\lambda )]  \notag \\
&=&\sum_{j=1}^{n}C_{r}[D(\lambda _{j})]  \notag \\
&=&\sum_{j=1}^{n}\sup_{\nu \geq 1}C_{r}[\rho (\nu I_{2},2\lambda _{j})]
\notag \\
&=&\sum_{j=1}^{n}\sup_{\nu \geq 1}[f(\frac{\nu -1}{2}+|\lambda _{j}|^{2})\
-f(\frac{\nu -1}{2})\ ]  \notag \\
&=&\sum_{j=1}^{n}f(|\lambda _{j}|^{2}),   \label{eq36}
\end{eqnarray}
the last step is because $f(x)$ is a concave function.

Besides $C_{r}$, we can also define the coherence measures for Gaussian channels via the coherence measures for Gaussian states based on the Bures metric and the Hellinger metric \cite{Illuminati-2016-arxiv} under Theorem 5.

We need to point out that there could be a divergence problem when we define a coherence measure for Gaussian channels under Theorem 5 since the supremum is taken over the unbounded set $\{\nu _{j}|\nu _{j}\geq
1\}_{j=1}^{n}$.

\section{Summary}

In this work, we proposed the definitions of incoherent Gaussian channel and
incoherent Gaussian superchannel, established a resource theory for
quantifying the coherence of Gaussian channels. We proposed two
representations for Gaussian superchannels and two representations for
incoherent Gaussian superchannels. We provided a way to construct a class of coherence measures for Gaussian channels via coherence measures for Gaussian states. It is worth emphasizing that
the definitions of incoherent Gaussian channels and incoherent Gaussian
superchannel in this work are all resource-nongenerating operations \cite%
{ChitambarGour-2019-RMP}. Two concrete examples are given to exemplify the
calculation of coherence measure $C_{r}$ for Gaussian channels.

\section*{ACKNOWLEDGMENTS}

This work is supported by the China Scholarship Council (CSC, No.
201806305050).

\section*{Appendix}

\setcounter{equation}{0} \renewcommand\theequation{A\arabic{equation}}

\subsection{Proof of Theorem 1}

Suppose $\phi (T,N,d)\in \mathcal{IGC}_{n}$, then for any $\rho
_{th}(V=\oplus _{j}\nu _{j}I_{2},d=0)$ we have $\phi (\rho _{th})\in
\mathcal{IGS}_{n}$. From Eqs. (\ref{eq8}, \ref{eq9}) it follows that $d=0$ and for any $\{\nu
_{j}|\nu _{j}\geq 1\}_{j=1}^{n}$, there exist $\{\nu _{j}^{\prime }|\nu
_{j}^{\prime }\geq 1\}_{j=1}^{n}$ such that
\begin{eqnarray}
T(\oplus _{j}\nu _{j}I_{2})T^{t}+N=\oplus _{j}\nu _{j}^{\prime }I_{2}. \label{eqA1}
\end{eqnarray}

Write $T$ as $n\times n$ block matrix $T=(T_{jk})_{jk}$ with each $T_{jk}$ a
$2\times 2$ real matrix located in $(2j-1,2j)$ rows and $(2k-1,2k)$ columns,
and $N=(N_{jk})_{jk}$ similarly. Then, Eq. (\ref{eqA1}) yields
\begin{eqnarray}
&&\sum_{l=1}^{n}\nu _{l}T_{jl}T_{kl}^{t}+N_{jk}=0,j\neq k, \label{eqA2} \\
&&\sum_{l=1}^{n}\nu _{l}T_{jl}T_{jl}^{t}+N_{jj}=\nu _{j}^{\prime }I_{2}, \label{eqA3}
\end{eqnarray}%
where $T_{kl}^{t}=(T_{kl})^{t}.$
Varying $\{\nu _{j}|\nu _{j}\geq 1\}_{j=1}^{n}$ in Eqs. (\ref{eqA2}, \ref{eqA3}) leads to
\begin{eqnarray}
&&T_{jl}T_{kl}^{t}=0,j\neq k, \label{eqA4} \\
&&N_{jk}=0,j\neq k, \label{eqA5} \\
&&T_{jl}T_{jl}^{t}=t_{jl}^{2}I_{2},t_{jl}\in R, \label{eqA6} \\
&&N_{jj}=\omega _{j}I_{2},\omega _{j}\in R. \label{eqA7}
\end{eqnarray}%
Eqs. (\ref{eqA4}, \ref{eqA6}) together imply that in $\{T_{jl}\}_{j=1}^{n}$ there is at most one
nonzero matrix, we denote this matrix by $t_{l}T_{l}$ with $t_{l}\in
R,T_{l}T_{l}^{t}=I_{2}.$

Taking these conditions into Eq. (\ref{eq10}) and using the fact that for any $2\times
2$ real matrix $M,$
\begin{eqnarray}
M\omega M^{t}=\omega \det M,
\end{eqnarray}
we get
\begin{eqnarray}
\omega _{j}I_{2}+i\omega (1-\sum_{k:r(k)=j}t_{k}^{2}\det T_{k})\geq
0,\forall \ j.
\end{eqnarray}
this evidently leads to Eq. (\ref{eq15}).

\subsection{Proof of Theorem 2}

For $\phi (T,N,d)\in \mathcal{GC}_{n}$, $\rho _{\phi }(V,d_{0})\in \mathcal{%
GS}_{2n}$, consider any $\Phi (X,Y_{0},\widetilde{d})\in \mathcal{GC}_{2n}$
with
\begin{eqnarray}
\widetilde{d}=\binom{\overline{d}}{0}.
\end{eqnarray}

Write
\begin{eqnarray}
X=\left(
\begin{array}{cc}
A & B \\
C & D%
\end{array}%
\right) ,Y=\left(
\begin{array}{cc}
A^{\prime } & B^{\prime } \\
(B^{\prime })^{t} & D^{\prime }%
\end{array}%
\right),
\end{eqnarray}
$A$, $B$, $C$, $D$, $A^{\prime }$, $B^{\prime }$, $D^{\prime }$, are all $%
2n\times 2n$ real matrices.

According to Eq. (\ref{eq9}) we have
\begin{eqnarray}
XVX^{t}+Y=\left(
\begin{array}{cc}
V_{11}^{\prime } & V_{12}^{\prime } \\
V_{21}^{\prime } & V_{22}^{\prime }%
\end{array}%
\right) +Y
\end{eqnarray}
with $V_{11}^{\prime },V_{12}^{\prime },V_{21}^{\prime },V_{22}^{\prime }$
are all $2n\times 2n$ real matrices as respectively
\begin{eqnarray}
&&(ATT^{t}A^{t}+BB^{t})ch2r+(B\Sigma _{n}T^{t}A^{t}+AT\Sigma _{n}B^{t})sh2r
\notag \\
&&+ANA^{t},\ \ \ \ \ \  \\
&&(ATT^{t}C^{t}+BD^{t})ch2r+(B\Sigma _{n}T^{t}C^{t}+AT\Sigma _{n}D^{t})sh2r
\notag \\
&&+ANC^{t},\ \ \ \ \ \  \\
&&(CTT^{t}A^{t}+DB^{t})ch2r+(D\Sigma _{n}T^{t}A^{t}+CT\Sigma _{n}B^{t})sh2r
\notag \\
&&+CNA^{t},\ \ \ \ \ \  \\
&&(CTT^{t}C^{t}+DD^{t})ch2r+(D\Sigma _{n}T^{t}C^{t}+CT\Sigma _{n}D^{t})sh2r
\notag \\
&&+CNC^{t}.\ \ \ \ \
\end{eqnarray}

If $XVX^{t}+Y$ has the form of Eq. (\ref{eq16}), noticing that r is varying, then
there must be
\begin{eqnarray}
&&ATT^{t}A^{t}+BB^{t}=T^{\prime }T^{\prime t}, \label{A17} \\
&&B\Sigma _{n}T^{t}A^{t}+AT\Sigma _{n}B^{t}=0,  \label{A18} \\
&&ATT^{t}C^{t}+BD^{t}=0,  \label{A19} \\
&&B\Sigma _{n}T^{t}C^{t}+AT\Sigma _{n}D^{t}=T^{\prime }\Sigma _{n},  \label{A20} \\
&&CTT^{t}C^{t}+DD^{t}=I_{2n},  \label{A21} \\
&&D\Sigma _{n}T^{t}C^{t}+CT\Sigma _{n}D^{t}=0,  \label{A22} \\
&&ANA^{t}+A^{\prime }=N^{\prime },  \label{A23} \\
&&ANC^{t}+B^{\prime }=0,  \label{A24} \\
&&CNC^{t}+D^{\prime }=0.  \label{A25}
\end{eqnarray}
Let $T=0$, $N+i\Omega \geq 0$, Eq. (\ref{A21}) and Eq. (\ref{A19}) yield
\begin{eqnarray}
&&DD^{t}=I_{2n}, \\
&&B=0.
\end{eqnarray}
Let $N=0$, $T=I_{n}$, it is easy to check that such $T$,
$N$ satisfy Eq. (\ref{eq10}). For such case, Eq. (\ref{A21}) yields
\begin{eqnarray}
C=0,
\end{eqnarray}
and then Eqs. (\ref{A24},\ref{A25}) lead to
\begin{eqnarray}
B^{\prime }=D^{\prime }=0.
\end{eqnarray}
Let $A^{\prime }=Y$, $D=O$, and notice that for $\Phi (X,Y_{0},\widetilde{d}%
)\in \mathcal{GC}_{2n}$, Eq. (\ref{eq8}) leads to Eq. (\ref{eq21}), Eq. (\ref{eq10}) leads to Eqs. (\ref{eq22}, \ref{eq23}), we then end this proof.

\subsection{Proof of Theorem 3}

For $\phi _{1}(T_{1},N_{1},d_{1})$, $\phi _{2}(T_{2},N_{2},d_{2})$, $\phi
_{3}(T_{3},N_{3},d_{3})\in \mathcal{GC}_{n}$, $\rho (V,d_{0})\in \mathcal{GS}%
_{n}$, denote $\rho (\widetilde{V},\widetilde{d_{0}})=\phi _{4}[\rho
(V,d_{0})]$, $\phi _{4}(T_{4},N_{4},d_{4})=\phi _{3}\circ \phi _{2}\circ
\phi _{1}$, then repeatedly using Eqs. (\ref{eq8}, \ref{eq9}) we get
\begin{eqnarray}
\widetilde{V}%
&=&T_{3}T_{2}T_{1}VT_{1}^{t}T_{2}^{t}T_{3}^{t}+T_{3}T_{2}N_{1}T_{2}^{t}T_{3}^{t}+T_{3}N_{2}T_{3}^{t}+N_{3},
\notag \\
\\
\widetilde{d_{0}}&=&T_{3}T_{2}T_{1}d_{0}+T_{3}T_{2}d_{1}+T_{3}d_{2}+d_{3}, \\
T_{4}&=&T_{3}T_{2}T_{1}, \\
N_{4}&=&T_{3}T_{2}N_{1}T_{2}^{t}T_{3}^{t}+T_{3}N_{2}T_{3}^{t}+N_{3}, \\
d_{4}&=&T_{3}T_{2}d_{1}+T_{3}d_{2}+d_{3}.
\end{eqnarray}
Now we can check that Eqs. (\ref{eq24}, \ref{eq25}) realize Eqs. (\ref{eq19}-\ref{eq21}).
Taking Eq. (\ref{eq25}) into Eq. (\ref{eq10}) we get Eq. (\ref{eq22}). Taking Eq. (\ref{eq24}) into Eq. (\ref{eq10}) we get
\begin{eqnarray}
i\Omega -i\Sigma _{n}O^{t}\Sigma _{n}\Omega \Sigma _{n}O\Sigma _{n}\geq 0.
\end{eqnarray}
Left multiply the left side of this equation by $O\Sigma _{n}$ and right multiply it by
$\Sigma _{n}O^{t}$, together with the fact that
\begin{eqnarray}
\Sigma _{n}\Omega \Sigma _{n}=-\Omega,
\end{eqnarray}
we will get Eq. (\ref{eq23}).

\subsection{Proof of Theorem 4}

Step 1. We prove $(1)\Rightarrow (2).$

Suppose Gaussian superchannel $\Phi (A,O,Y,\overline{d})$ is incoherent, then for any $\phi (T,N,d)\in \mathcal{IGC}_{n}$, we have $\Phi \lbrack \phi (T,N,d)]=\psi (T^{\prime},N^{\prime },d^{\prime })\in \mathcal{IGC}_{n}$. From Theorem 1 and Theorem 2,
similarly to the proof of Theorem 1, we can get $\overline{d}=0$, $A\in
\mathcal{T}_{n}$, $Y=\oplus _{j}\eta _{j}I_{2}$, $\eta _{j}\in R$. We need
to prove $O\in \mathcal{T}_{n}$. Let
\begin{eqnarray}
O^{\prime }=\Sigma _{n}O^{t}\Sigma _{n},
\end{eqnarray}
then $T\in \mathcal{T}_{n}$, $A\in \mathcal{T}_{n}$, and
\begin{eqnarray}
T^{\prime }=ATO^{\prime }\in \mathcal{T}_{n}.
\end{eqnarray}
Note that $\mathcal{T}_{n}$ is closed under multiplication, then $AT\in
\mathcal{T}_{n}.$

If $A=0$, then $T^{\prime }=ATO^{\prime }=0,$ we let $O=O^{\prime }=I_{n}\in \mathcal{T}_{n}.$

Suppose $\mathcal{T}_{n}\mathcal{\ni }A=\{s_{j}A_{j}\}_{j}\neq 0,$ then there exists at least one $j_{0}$
such that $s_{j_{0}}\neq 0$, $A_{j_{0}}A_{j_{0}}^{t}=I_{2}.$ Let $%
T=\{t_{j}T_{j}\}_{j}$ with $T_{j}T_{j}^{t}=I_{2}$, $r(T_{j})=j_{0}$ $\forall
$ $j$, then $\mathcal{T}_{n}\mathcal{\ni }AT=\{s_{j_{0}}t_{j}A_{j_{0}}T_{j}%
\}_{j}$ with $r(A_{j_{0}}T_{j})=r(A_{j_{0}})$ $\forall $ $j$. Now write $%
O^{\prime }$ as $n\times n$ block matrix $O^{\prime }=(O_{jk}^{\prime
})_{jk} $ with each $O_{jk}^{\prime }$ a $2\times 2$ real matrix located in $%
(2j-1,2j)$ rows and $(2k-1,2k)$ columns, then
\begin{eqnarray}
\mathcal{T}_{n}\mathcal{\ni }T^{\prime }=ATO^{\prime
}=\{s_{j_{0}}A_{j_{0}}\sum_{j}t_{j}T_{j}O_{jk}^{\prime }\}_{k}.
\end{eqnarray}
Varying $t_{j}$, $T_{j}$, for all $j$, and using the facts of
Lemma 1 and lemma 2 below, we can get $O^{\prime }$ $\in \mathcal{T}_{n}$ and further $%
O=\Sigma _{n}(O^{\prime })^{t}\Sigma _{n}\in \mathcal{T}_{n}.$

\textbf{Lemma 1.} Let $B_{1},B_{2}$ be two fixed $2\times 2$ real matrices.
If for any $t_{1}$, $t_{2}\in R,$ there exists $t\in R$ such that
\begin{eqnarray}
(t_{1}B_{1}+t_{2}B_{2})(t_{1}B_{1}+t_{2}B_{2})^{t}=tI_{2},  \label{A40} 
\end{eqnarray}
then there exist $s_{1},s_{2},s_{3}\in R$ such that
\begin{eqnarray}
&&B_{1}B_{1}^{t}=s_{1}^{2}I_{2}, \label{A41} \\
&&B_{2}B_{2}^{t}=s_{2}^{2}I_{2},  \label{A42}  \\
&&B_{1}B_{2}^{t}+B_{2}B_{1}^{t}=s_{3}I_{2}.  \label{A43} 
\end{eqnarray}

Expand Eq. (\ref{A40}) and vary $t_{1}$, $t_{2}\in R$ we will get Lemma 1. Note that there are $2\times 2$ real matrices $B_{1},B_{2}$ satisfying Eqs.
(\ref{A41}, \ref{A42}) but not satisfying Eq. (\ref{A43}), for example $B_{1}=I_{2}$, $B_{2}=\frac{1}{%
\sqrt{2}}\left(
\begin{array}{cc}
1 & 1 \\
1 & -1%
\end{array}%
\right) $.

\textbf{Lemma 2.} Let $B_{2}$ be a fixed $2\times 2$ real matrix.
If for any $t_{1}$, $t_{2}\in R$ and any $2\times 2$ real matrix $B_{1}\neq 0$, there exists $t\in R$ such that
\begin{eqnarray}
(t_{1}B_{1}+t_{2}B_{2})(t_{1}B_{1}+t_{2}B_{2})^{t}=tI_{2},  \label{A44} 
\end{eqnarray}
then $B_{2}=0.$

Proof of Lemma 2. Any $2\times 2$ real orthogonal matrix has one of the forms 
 \begin{eqnarray}
Q_{1}(\theta)=\left(
\begin{array}{cc}
cos\theta & -sin\theta \\
sin\theta & cos\theta%
\end{array}
\right), \\
Q_{2}(\theta)=\left(
\begin{array}{cc}
cos\theta & sin\theta \\
sin\theta & -cos\theta%
\end{array}
\right),
\end{eqnarray}
and 
\begin{eqnarray}
Q_{1}(\theta _{1})Q_{1}(\theta _{2})&=&Q_{1}(\theta _{1}+\theta  _{2}),  \label{A47} \\
Q_{2}(\theta _{1})Q_{2}(\theta _{2})&=&Q_{1}(\theta _{1}-\theta  _{2}),  \label{A48} \\
Q_{1}(\theta _{1})Q_{2}(\theta _{2})&=&Q_{2}(\theta _{1}+\theta  _{2}),  \label{A49} \\
Q_{2}(\theta _{1})Q_{1}(\theta _{2})&=&Q_{2}(-\theta _{1}+\theta  _{2}),  \label{A50}
\end{eqnarray}
where $\theta, \theta _{1}, \theta _{2}\in R.$
With the fact of lemma 1, taking (\ref{A47}-\ref{A50}) into Eq. (\ref{A43}) and varying $B_{1}$, we will get Lemma 2.
 
Step 2. From Theorem 3 we can get $(2)\Rightarrow (3).$

Step 3. For $\chi _{1}$, $\phi $, $\chi _{2}\in \mathcal{IGC}_{n}$, we
evidently have $\Phi (\phi )=\chi _{2}\circ \phi \circ \chi _{1}\in \mathcal{%
IGC}_{n},$ hence $(3)\Rightarrow (1)$.

\bibliographystyle{apsrev4-1}
\bibliography{CoherenceofGaussianchannels}

\begin{thebibliography}{25}%
\makeatletter
\providecommand \@ifxundefined [1]{%
 \@ifx{#1\undefined}
}%
\providecommand \@ifnum [1]{%
 \ifnum #1\expandafter \@firstoftwo
 \else \expandafter \@secondoftwo
 \fi
}%
\providecommand \@ifx [1]{%
 \ifx #1\expandafter \@firstoftwo
 \else \expandafter \@secondoftwo
 \fi
}%
\providecommand \natexlab [1]{#1}%
\providecommand \enquote  [1]{``#1''}%
\providecommand \bibnamefont  [1]{#1}%
\providecommand \bibfnamefont [1]{#1}%
\providecommand \citenamefont [1]{#1}%
\providecommand \href@noop [0]{\@secondoftwo}%
\providecommand \href [0]{\begingroup \@sanitize@url \@href}%
\providecommand \@href[1]{\@@startlink{#1}\@@href}%
\providecommand \@@href[1]{\endgroup#1\@@endlink}%
\providecommand \@sanitize@url [0]{\catcode `\\12\catcode `\$12\catcode
  `\&12\catcode `\#12\catcode `\^12\catcode `\_12\catcode `\%12\relax}%
\providecommand \@@startlink[1]{}%
\providecommand \@@endlink[0]{}%
\providecommand \url  [0]{\begingroup\@sanitize@url \@url }%
\providecommand \@url [1]{\endgroup\@href {#1}{\urlprefix }}%
\providecommand \urlprefix  [0]{URL }%
\providecommand \Eprint [0]{\href }%
\providecommand \doibase [0]{http://dx.doi.org/}%
\providecommand \selectlanguage [0]{\@gobble}%
\providecommand \bibinfo  [0]{\@secondoftwo}%
\providecommand \bibfield  [0]{\@secondoftwo}%
\providecommand \translation [1]{[#1]}%
\providecommand \BibitemOpen [0]{}%
\providecommand \bibitemStop [0]{}%
\providecommand \bibitemNoStop [0]{.\EOS\space}%
\providecommand \EOS [0]{\spacefactor3000\relax}%
\providecommand \BibitemShut  [1]{\csname bibitem#1\endcsname}%
\let\auto@bib@innerbib\@empty
\bibitem [{\citenamefont {Streltsov}\ \emph {et~al.}(2017)\citenamefont
  {Streltsov}, \citenamefont {Adesso},\ and\ \citenamefont
  {Plenio}}]{Plenio-2016-RMP}%
  \BibitemOpen
  \bibfield  {author} {\bibinfo {author} {\bibfnamefont {A.}~\bibnamefont
  {Streltsov}}, \bibinfo {author} {\bibfnamefont {G.}~\bibnamefont {Adesso}}, \
  and\ \bibinfo {author} {\bibfnamefont {M.~B.}\ \bibnamefont {Plenio}},\
  }\href {\doibase 10.1103/RevModPhys.89.041003} {\bibfield  {journal}
  {\bibinfo  {journal} {Rev. Mod. Phys.}\ }\textbf {\bibinfo {volume} {89}},\
  \bibinfo {pages} {041003} (\bibinfo {year} {2017})}\BibitemShut {NoStop}%
\bibitem [{\citenamefont {Hu}\ \emph {et~al.}(2018)\citenamefont {Hu},
  \citenamefont {Hu}, \citenamefont {Wang}, \citenamefont {Peng}, \citenamefont
  {Zhang},\ and\ \citenamefont {Fan}}]{Hu2018-PhysicsReports}%
  \BibitemOpen
  \bibfield  {author} {\bibinfo {author} {\bibfnamefont {M.-L.}\ \bibnamefont
  {Hu}}, \bibinfo {author} {\bibfnamefont {X.}~\bibnamefont {Hu}}, \bibinfo
  {author} {\bibfnamefont {J.}~\bibnamefont {Wang}}, \bibinfo {author}
  {\bibfnamefont {Y.}~\bibnamefont {Peng}}, \bibinfo {author} {\bibfnamefont
  {Y.-R.}\ \bibnamefont {Zhang}}, \ and\ \bibinfo {author} {\bibfnamefont
  {H.}~\bibnamefont {Fan}},\ }\href {\doibase
  https://doi.org/10.1016/j.physrep.2018.07.004} {\bibfield  {journal}
  {\bibinfo  {journal} {Physics Reports}\ }\textbf {\bibinfo {volume}
  {762-764}},\ \bibinfo {pages} {1 } (\bibinfo {year} {2018})}\BibitemShut
  {NoStop}%
\bibitem [{\citenamefont {{Gour}}(2019)}]{Gour-2019-IEEE}%
  \BibitemOpen
  \bibfield  {author} {\bibinfo {author} {\bibfnamefont {G.}~\bibnamefont
  {{Gour}}},\ }\href {\doibase 10.1109/TIT.2019.2907989} {\bibfield  {journal}
  {\bibinfo  {journal} {IEEE Transactions on Information Theory}\ ,\ \bibinfo
  {pages} {1}} (\bibinfo {year} {2019})}\BibitemShut {NoStop}%
\bibitem [{\citenamefont {{Devetak}}\ \emph {et~al.}(2008)\citenamefont
  {{Devetak}}, \citenamefont {{Harrow}},\ and\ \citenamefont
  {{Winter}}}]{Winter-IEEE-2008}%
  \BibitemOpen
  \bibfield  {author} {\bibinfo {author} {\bibfnamefont {I.}~\bibnamefont
  {{Devetak}}}, \bibinfo {author} {\bibfnamefont {A.~W.}\ \bibnamefont
  {{Harrow}}}, \ and\ \bibinfo {author} {\bibfnamefont {A.~J.}\ \bibnamefont
  {{Winter}}},\ }\href {\doibase 10.1109/TIT.2008.928980} {\bibfield  {journal}
  {\bibinfo  {journal} {IEEE Transactions on Information Theory}\ }\textbf
  {\bibinfo {volume} {54}},\ \bibinfo {pages} {4587} (\bibinfo {year}
  {2008})}\BibitemShut {NoStop}%
\bibitem [{\citenamefont {Ben~Dana}\ \emph {et~al.}(2017)\citenamefont
  {Ben~Dana}, \citenamefont {Garc\'{\i}a~D\'{\i}az}, \citenamefont {Mejatty},\
  and\ \citenamefont {Winter}}]{Winter-2017-PRA}%
  \BibitemOpen
  \bibfield  {author} {\bibinfo {author} {\bibfnamefont {K.}~\bibnamefont
  {Ben~Dana}}, \bibinfo {author} {\bibfnamefont {M.}~\bibnamefont
  {Garc\'{\i}a~D\'{\i}az}}, \bibinfo {author} {\bibfnamefont {M.}~\bibnamefont
  {Mejatty}}, \ and\ \bibinfo {author} {\bibfnamefont {A.}~\bibnamefont
  {Winter}},\ }\href {\doibase 10.1103/PhysRevA.95.062327} {\bibfield
  {journal} {\bibinfo  {journal} {Phys. Rev. A}\ }\textbf {\bibinfo {volume}
  {95}},\ \bibinfo {pages} {062327} (\bibinfo {year} {2017})}\BibitemShut
  {NoStop}%
\bibitem [{\citenamefont {Korzekwa}\ \emph {et~al.}(2018)\citenamefont
  {Korzekwa}, \citenamefont {Czach{\'{o}}rski}, \citenamefont {Pucha{\l}a},\
  and\ \citenamefont {{\.{Z}}yczkowski}}]{Korzekwa-2018-NJP}%
  \BibitemOpen
  \bibfield  {author} {\bibinfo {author} {\bibfnamefont {K.}~\bibnamefont
  {Korzekwa}}, \bibinfo {author} {\bibfnamefont {S.}~\bibnamefont
  {Czach{\'{o}}rski}}, \bibinfo {author} {\bibfnamefont {Z.}~\bibnamefont
  {Pucha{\l}a}}, \ and\ \bibinfo {author} {\bibfnamefont {K.}~\bibnamefont
  {{\.{Z}}yczkowski}},\ }\href {\doibase 10.1088/1367-2630/aaaff3} {\bibfield
  {journal} {\bibinfo  {journal} {New Journal of Physics}\ }\textbf {\bibinfo
  {volume} {20}},\ \bibinfo {pages} {043028} (\bibinfo {year}
  {2018})}\BibitemShut {NoStop}%
\bibitem [{\citenamefont {Datta}\ \emph {et~al.}(2018)\citenamefont {Datta},
  \citenamefont {Sazim}, \citenamefont {Pati},\ and\ \citenamefont
  {Agrawal}}]{Datta-2018-AOP}%
  \BibitemOpen
  \bibfield  {author} {\bibinfo {author} {\bibfnamefont {C.}~\bibnamefont
  {Datta}}, \bibinfo {author} {\bibfnamefont {S.}~\bibnamefont {Sazim}},
  \bibinfo {author} {\bibfnamefont {A.~K.}\ \bibnamefont {Pati}}, \ and\
  \bibinfo {author} {\bibfnamefont {P.}~\bibnamefont {Agrawal}},\ }\href
  {\doibase https://doi.org/10.1016/j.aop.2018.08.014} {\bibfield  {journal}
  {\bibinfo  {journal} {Annals of Physics}\ }\textbf {\bibinfo {volume}
  {397}},\ \bibinfo {pages} {243 } (\bibinfo {year} {2018})}\BibitemShut
  {NoStop}%
\bibitem [{\citenamefont {Theurer}\ \emph {et~al.}(2019)\citenamefont
  {Theurer}, \citenamefont {Egloff}, \citenamefont {Zhang},\ and\ \citenamefont
  {Plenio}}]{Plenio-2019-PRL}%
  \BibitemOpen
  \bibfield  {author} {\bibinfo {author} {\bibfnamefont {T.}~\bibnamefont
  {Theurer}}, \bibinfo {author} {\bibfnamefont {D.}~\bibnamefont {Egloff}},
  \bibinfo {author} {\bibfnamefont {L.}~\bibnamefont {Zhang}}, \ and\ \bibinfo
  {author} {\bibfnamefont {M.~B.}\ \bibnamefont {Plenio}},\ }\href {\doibase
  10.1103/PhysRevLett.122.190405} {\bibfield  {journal} {\bibinfo  {journal}
  {Phys. Rev. Lett.}\ }\textbf {\bibinfo {volume} {122}},\ \bibinfo {pages}
  {190405} (\bibinfo {year} {2019})}\BibitemShut {NoStop}%
\bibitem [{\citenamefont {Liu}\ and\ \citenamefont
  {Yuan}(2019)}]{LiuYuan-2019}%
  \BibitemOpen
  \bibfield  {author} {\bibinfo {author} {\bibfnamefont {Y.}~\bibnamefont
  {Liu}}\ and\ \bibinfo {author} {\bibfnamefont {X.}~\bibnamefont {Yuan}},\
  }\href {https://arxiv.org/abs/1904.02680} {\bibfield  {journal} {\bibinfo
  {journal} {arXiv:1904.02680}\ } (\bibinfo {year} {2019})}\BibitemShut
  {NoStop}%
\bibitem [{\citenamefont {Liu}\ and\ \citenamefont
  {Winter}(2019)}]{LiuWinter-2019}%
  \BibitemOpen
  \bibfield  {author} {\bibinfo {author} {\bibfnamefont {Z.-W.}\ \bibnamefont
  {Liu}}\ and\ \bibinfo {author} {\bibfnamefont {A.}~\bibnamefont {Winter}},\
  }\href {https://arxiv.org/abs/1904.04201} {\bibfield  {journal} {\bibinfo
  {journal} {arXiv:1904.04201}\ } (\bibinfo {year} {2019})}\BibitemShut
  {NoStop}%
\bibitem [{\citenamefont {Xu}(2019)}]{Xu-2019-arXiv}%
  \BibitemOpen
  \bibfield  {author} {\bibinfo {author} {\bibfnamefont {J.}~\bibnamefont
  {Xu}},\ }\href {https://arxiv.org/abs/1907.07289} {\bibfield  {journal}
  {\bibinfo  {journal} {arXiv:1907.07289}\ } (\bibinfo {year}
  {2019})}\BibitemShut {NoStop}%
\bibitem [{\citenamefont {Baumgratz}\ \emph {et~al.}(2014)\citenamefont
  {Baumgratz}, \citenamefont {Cramer},\ and\ \citenamefont
  {Plenio}}]{BCP-2014-PRL}%
  \BibitemOpen
  \bibfield  {author} {\bibinfo {author} {\bibfnamefont {T.}~\bibnamefont
  {Baumgratz}}, \bibinfo {author} {\bibfnamefont {M.}~\bibnamefont {Cramer}}, \
  and\ \bibinfo {author} {\bibfnamefont {M.~B.}\ \bibnamefont {Plenio}},\
  }\href {\doibase 10.1103/PhysRevLett.113.140401} {\bibfield  {journal}
  {\bibinfo  {journal} {Phys. Rev. Lett.}\ }\textbf {\bibinfo {volume} {113}},\
  \bibinfo {pages} {140401} (\bibinfo {year} {2014})}\BibitemShut {NoStop}%
\bibitem [{\citenamefont {Weedbrook}\ \emph {et~al.}(2012)\citenamefont
  {Weedbrook}, \citenamefont {Pirandola}, \citenamefont {Garc\'{\i}a-Patr\'on},
  \citenamefont {Cerf}, \citenamefont {Ralph}, \citenamefont {Shapiro},\ and\
  \citenamefont {Lloyd}}]{Lloyd-2012-RMP}%
  \BibitemOpen
  \bibfield  {author} {\bibinfo {author} {\bibfnamefont {C.}~\bibnamefont
  {Weedbrook}}, \bibinfo {author} {\bibfnamefont {S.}~\bibnamefont
  {Pirandola}}, \bibinfo {author} {\bibfnamefont {R.}~\bibnamefont
  {Garc\'{\i}a-Patr\'on}}, \bibinfo {author} {\bibfnamefont {N.~J.}\
  \bibnamefont {Cerf}}, \bibinfo {author} {\bibfnamefont {T.~C.}\ \bibnamefont
  {Ralph}}, \bibinfo {author} {\bibfnamefont {J.~H.}\ \bibnamefont {Shapiro}},
  \ and\ \bibinfo {author} {\bibfnamefont {S.}~\bibnamefont {Lloyd}},\ }\href
  {\doibase 10.1103/RevModPhys.84.621} {\bibfield  {journal} {\bibinfo
  {journal} {Rev. Mod. Phys.}\ }\textbf {\bibinfo {volume} {84}},\ \bibinfo
  {pages} {621} (\bibinfo {year} {2012})}\BibitemShut {NoStop}%
\bibitem [{\citenamefont {Serafini}(2017)}]{Serafini-2017-book}%
  \BibitemOpen
  \bibfield  {author} {\bibinfo {author} {\bibfnamefont {A.}~\bibnamefont
  {Serafini}},\ }\href {\doibase 10.1201/9781315118727} {\emph {\bibinfo
  {title} {Quantum Continuous Variables}}}\ (\bibinfo  {publisher} {CRC
  Press},\ \bibinfo {address} {Boca Raton},\ \bibinfo {year}
  {2017})\BibitemShut {NoStop}%
\bibitem [{\citenamefont {Peng}\ \emph {et~al.}(2016)\citenamefont {Peng},
  \citenamefont {Jiang},\ and\ \citenamefont {Fan}}]{Fan-2016-PRA}%
  \BibitemOpen
  \bibfield  {author} {\bibinfo {author} {\bibfnamefont {Y.}~\bibnamefont
  {Peng}}, \bibinfo {author} {\bibfnamefont {Y.}~\bibnamefont {Jiang}}, \ and\
  \bibinfo {author} {\bibfnamefont {H.}~\bibnamefont {Fan}},\ }\href {\doibase
  10.1103/PhysRevA.93.032326} {\bibfield  {journal} {\bibinfo  {journal} {Phys.
  Rev. A}\ }\textbf {\bibinfo {volume} {93}},\ \bibinfo {pages} {032326}
  (\bibinfo {year} {2016})}\BibitemShut {NoStop}%
\bibitem [{\citenamefont {Xu}(2016)}]{Xu-2016-PRA}%
  \BibitemOpen
  \bibfield  {author} {\bibinfo {author} {\bibfnamefont {J.}~\bibnamefont
  {Xu}},\ }\href {\doibase 10.1103/PhysRevA.93.032111} {\bibfield  {journal}
  {\bibinfo  {journal} {Phys. Rev. A}\ }\textbf {\bibinfo {volume} {93}},\
  \bibinfo {pages} {032111} (\bibinfo {year} {2016})}\BibitemShut {NoStop}%
\bibitem [{\citenamefont {Zheng}\ \emph {et~al.}(2016)\citenamefont {Zheng},
  \citenamefont {Xu}, \citenamefont {Yao},\ and\ \citenamefont
  {Li}}]{Zheng-2016-PRA}%
  \BibitemOpen
  \bibfield  {author} {\bibinfo {author} {\bibfnamefont {Q.}~\bibnamefont
  {Zheng}}, \bibinfo {author} {\bibfnamefont {J.}~\bibnamefont {Xu}}, \bibinfo
  {author} {\bibfnamefont {Y.}~\bibnamefont {Yao}}, \ and\ \bibinfo {author}
  {\bibfnamefont {Y.}~\bibnamefont {Li}},\ }\href {\doibase
  10.1103/PhysRevA.94.052314} {\bibfield  {journal} {\bibinfo  {journal} {Phys.
  Rev. A}\ }\textbf {\bibinfo {volume} {94}},\ \bibinfo {pages} {052314}
  (\bibinfo {year} {2016})}\BibitemShut {NoStop}%
\bibitem [{\citenamefont {Buono}\ \emph {et~al.}(2016)\citenamefont {Buono},
  \citenamefont {Nocerino}, \citenamefont {Petrillo}, \citenamefont {Torre},
  \citenamefont {Zonzo},\ and\ \citenamefont
  {Illuminati}}]{Illuminati-2016-arxiv}%
  \BibitemOpen
  \bibfield  {author} {\bibinfo {author} {\bibfnamefont {D.}~\bibnamefont
  {Buono}}, \bibinfo {author} {\bibfnamefont {G.}~\bibnamefont {Nocerino}},
  \bibinfo {author} {\bibfnamefont {G.}~\bibnamefont {Petrillo}}, \bibinfo
  {author} {\bibfnamefont {G.}~\bibnamefont {Torre}}, \bibinfo {author}
  {\bibfnamefont {G.}~\bibnamefont {Zonzo}}, \ and\ \bibinfo {author}
  {\bibfnamefont {F.}~\bibnamefont {Illuminati}},\ }\href
  {https://arxiv.org/abs/1609.00913} {\bibfield  {journal} {\bibinfo  {journal}
  {arXiv:1609.00913}\ } (\bibinfo {year} {2016})}\BibitemShut {NoStop}%
\bibitem [{\citenamefont {Tan}\ \emph {et~al.}(2017)\citenamefont {Tan},
  \citenamefont {Volkoff}, \citenamefont {Kwon},\ and\ \citenamefont
  {Jeong}}]{Jeong-2017-PRL}%
  \BibitemOpen
  \bibfield  {author} {\bibinfo {author} {\bibfnamefont {K.~C.}\ \bibnamefont
  {Tan}}, \bibinfo {author} {\bibfnamefont {T.}~\bibnamefont {Volkoff}},
  \bibinfo {author} {\bibfnamefont {H.}~\bibnamefont {Kwon}}, \ and\ \bibinfo
  {author} {\bibfnamefont {H.}~\bibnamefont {Jeong}},\ }\href {\doibase
  10.1103/PhysRevLett.119.190405} {\bibfield  {journal} {\bibinfo  {journal}
  {Phys. Rev. Lett.}\ }\textbf {\bibinfo {volume} {119}},\ \bibinfo {pages}
  {190405} (\bibinfo {year} {2017})}\BibitemShut {NoStop}%
\bibitem [{\citenamefont {Albarelli}\ \emph {et~al.}(2017)\citenamefont
  {Albarelli}, \citenamefont {Genoni},\ and\ \citenamefont
  {Paris}}]{Paris-2017-PRA}%
  \BibitemOpen
  \bibfield  {author} {\bibinfo {author} {\bibfnamefont {F.}~\bibnamefont
  {Albarelli}}, \bibinfo {author} {\bibfnamefont {M.~G.}\ \bibnamefont
  {Genoni}}, \ and\ \bibinfo {author} {\bibfnamefont {M.~G.~A.}\ \bibnamefont
  {Paris}},\ }\href {\doibase 10.1103/PhysRevA.96.012337} {\bibfield  {journal}
  {\bibinfo  {journal} {Phys. Rev. A}\ }\textbf {\bibinfo {volume} {96}},\
  \bibinfo {pages} {012337} (\bibinfo {year} {2017})}\BibitemShut {NoStop}%
\bibitem [{\citenamefont {Chitambar}\ and\ \citenamefont
  {Gour}(2019)}]{ChitambarGour-2019-RMP}%
  \BibitemOpen
  \bibfield  {author} {\bibinfo {author} {\bibfnamefont {E.}~\bibnamefont
  {Chitambar}}\ and\ \bibinfo {author} {\bibfnamefont {G.}~\bibnamefont
  {Gour}},\ }\href {\doibase 10.1103/RevModPhys.91.025001} {\bibfield
  {journal} {\bibinfo  {journal} {Rev. Mod. Phys.}\ }\textbf {\bibinfo {volume}
  {91}},\ \bibinfo {pages} {025001} (\bibinfo {year} {2019})}\BibitemShut
  {NoStop}%
\bibitem [{\citenamefont {Simon}\ \emph {et~al.}(1994)\citenamefont {Simon},
  \citenamefont {Mukunda},\ and\ \citenamefont {Dutta}}]{Simon-1994-PRA}%
  \BibitemOpen
  \bibfield  {author} {\bibinfo {author} {\bibfnamefont {R.}~\bibnamefont
  {Simon}}, \bibinfo {author} {\bibfnamefont {N.}~\bibnamefont {Mukunda}}, \
  and\ \bibinfo {author} {\bibfnamefont {B.}~\bibnamefont {Dutta}},\ }\href
  {\doibase 10.1103/PhysRevA.49.1567} {\bibfield  {journal} {\bibinfo
  {journal} {Phys. Rev. A}\ }\textbf {\bibinfo {volume} {49}},\ \bibinfo
  {pages} {1567} (\bibinfo {year} {1994})}\BibitemShut {NoStop}%
\bibitem [{\citenamefont {Nielsen}\ and\ \citenamefont
  {Chuang}(2000)}]{Nielsen-2000-book}%
  \BibitemOpen
  \bibfield  {author} {\bibinfo {author} {\bibfnamefont {M.~A.}\ \bibnamefont
  {Nielsen}}\ and\ \bibinfo {author} {\bibfnamefont {I.~L.}\ \bibnamefont
  {Chuang}},\ }\href@noop {} {\emph {\bibinfo {title} {Quantum Computation and
  Quantum Information}}}\ (\bibinfo  {publisher} {Cambridge University Press},\
  \bibinfo {address} {Cambridge},\ \bibinfo {year} {2000})\BibitemShut
  {NoStop}%
\bibitem [{\citenamefont {Holevo}\ \emph {et~al.}(1999)\citenamefont {Holevo},
  \citenamefont {Sohma},\ and\ \citenamefont {Hirota}}]{Holevo-1999-PRA}%
  \BibitemOpen
  \bibfield  {author} {\bibinfo {author} {\bibfnamefont {A.~S.}\ \bibnamefont
  {Holevo}}, \bibinfo {author} {\bibfnamefont {M.}~\bibnamefont {Sohma}}, \
  and\ \bibinfo {author} {\bibfnamefont {O.}~\bibnamefont {Hirota}},\ }\href
  {\doibase 10.1103/PhysRevA.59.1820} {\bibfield  {journal} {\bibinfo
  {journal} {Phys. Rev. A}\ }\textbf {\bibinfo {volume} {59}},\ \bibinfo
  {pages} {1820} (\bibinfo {year} {1999})}\BibitemShut {NoStop}%
\bibitem [{\citenamefont {Chiribella}\ \emph {et~al.}(2008)\citenamefont
  {Chiribella}, \citenamefont {DAriano},\ and\ \citenamefont
  {Perinotti}}]{Chiribella-2008-EPL}%
  \BibitemOpen
  \bibfield  {author} {\bibinfo {author} {\bibfnamefont {G.}~\bibnamefont
  {Chiribella}}, \bibinfo {author} {\bibfnamefont {G.~M.}\ \bibnamefont
  {DAriano}}, \ and\ \bibinfo {author} {\bibfnamefont {P.}~\bibnamefont
  {Perinotti}},\ }\href {\doibase 10.1209/0295-5075/83/30004} {\bibfield
  {journal} {\bibinfo  {journal} {{EPL} (Europhysics Letters)}\ }\textbf
  {\bibinfo {volume} {83}},\ \bibinfo {pages} {30004} (\bibinfo {year}
  {2008})}\BibitemShut {NoStop}%
\end{thebibliography}%

\end{document}